\documentclass[aps,prl,amsmath,amssymb,floatfix,twocolumn,amsmath,superscriptaddress,twocolumn,nofootinbib,tighten,letterpaper]{revtex4-1}
\usepackage{multirow}
\usepackage{bbold}

\usepackage{color}
\usepackage{mathrsfs}
\usepackage{hyperref}
\usepackage[normalem]{ulem}
\usepackage{bm}

\usepackage[caption=false]{subfig} 

\usepackage{amsfonts, relsize, color}
\usepackage{graphics}
\usepackage{graphicx}

\usepackage{hyperref}
\usepackage{color}
\usepackage{comment}

\renewcommand\vec[1]{\ensuremath\boldsymbol{#1}} 

\begin{document}
\title{Metals, fractional metals, and superconductivity in rhombohedral trilayer graphene}

\author{Andr$\acute{\mbox{a}}$s L. Szab$\acute{\mbox{o}}$}
\affiliation{Max-Planck-Institut f\"{u}r Physik komplexer Systeme, N\"{o}thnitzer Str. 38, 01187 Dresden, Germany}

\author{Bitan Roy}\thanks{Corresponding author: bitan.roy@lehigh.edu}
\affiliation{Department of Physics, Lehigh University, Bethlehem, Pennsylvania, 18015, USA}

\date{\today}
\begin{abstract}
Combining mean-field and renormalization group analyses, here we unveil the nature of recently observed superconductivity and parent metallic states in chemically doped rhombohedral trilayer graphene, subject to external electric displacement fields ($D$) [H. Zhou, \emph{et al.}, Nature (London) {\bf 598}, 434 (2021)]. We argue that close to the charge neutrality, on site Hubbard repulsion favors layer antiferromagnet, which when combined with the $D$-field induced layer polarization, produces a spin-polarized, but valley-unpolarized half-metal, conducive to the nucleation of spin-triplet $f$-wave pairing (SC2). At larger doping valence bond order emerges as a prominent candidate for isospin coherent paramagent, boosting condensation of spin-singlet Cooper pairs in the $s$-wave channel (SC1), manifesting a ``selection rule" among competing orders. Responses of these paired states to displacement and in-plane magnetic fields show qualitative similarities with experimental observation. With the onset of the quantum anomalous Hall order, the valley degeneracy of half-metal gets lifted, forming a quarter-metal at lower doping [H. Zhou, \emph{et al.}, Nature (London) {\bf 598}, 429 (2021)].         
\end{abstract}

\maketitle

\emph{Introduction}.~Carbon based atom-thick stacked layers of honeycomb membrane open up a rich landscape harboring peculiar band dispersion of gapless chiral quasiparticles~\cite{katsnelson:book, graphene:RMP}. As such, Bernal bilayer graphene and rhombohedral trilayer graphene (RTLG) respectively accommodate bi-quadratic~\cite{BLG:band} and bi-cubic~\cite{RTLG:band} band touchings at two independent corners of the hexagonal Brillouin zone, giving rise to SU(2) valley or isospin degrees of freedom. Twist by a relatively small, so-called magic angle ($\sim 1^{\circ}$) between two honeycomb layers produces nearly flat bands of massless Dirac fermions~\cite{TBLGband:1, TBLGband:2, TBLGband:3, TBLGband:4, TBLGband:5}, where superconductivity has been observed~\cite{TBLGSC:1, TBLGSC:2, TBLGSC:3, TBLGSC:4}. More recently, superconductivity in RTLG has been reported, when it is chemically doped and subject to external electric displacement field ($D$) in the stacking direction~\cite{young:RTLGSC}. Culmination of these experimental achievements places us at the dawn of the carbon age of superconductivity. This Letter unfolds the nature of the superconducting orders and their parent (half)metallic states in RTLG.

\begin{figure}[t!]
\includegraphics[width=0.75\linewidth]{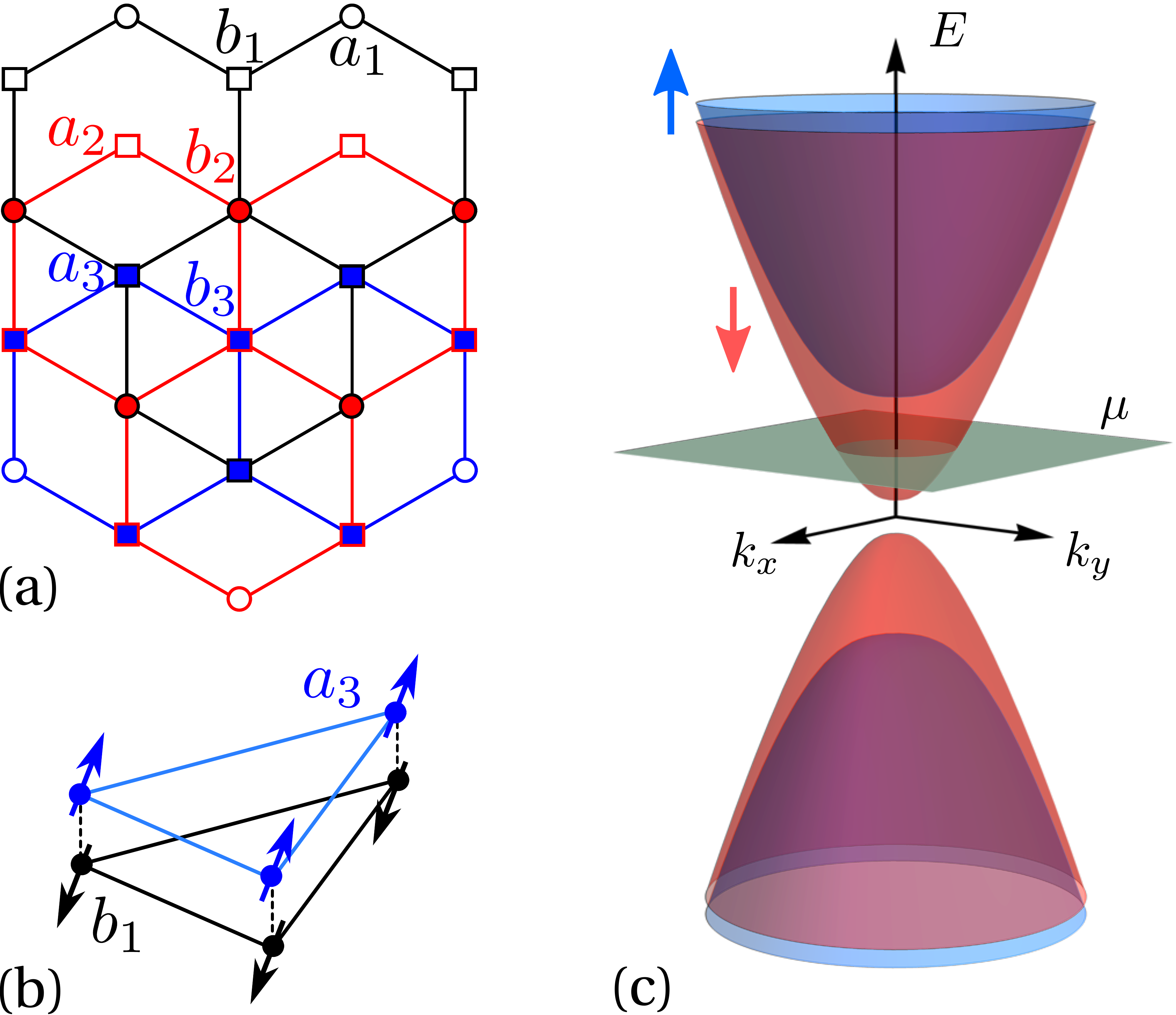}
\caption{(a) Top view of RTLG. The subscript $i=1,2,3$ denote the layer index of the sites. Each $a_1$ and $b_2$ sites, as well as each $a_2$ and $b_3$ sites overlap. Eigenstates of the high energy split-off bands reside dominantly on these four dimer sites. Sites $b_1$ and $a_3$ form an effective bipartite lattice and participate in the low-energy description of RTLG, featuring bi-cubic band touchings [Eq.~(\ref{eq:hamilRTLG})]. (b) Layer antiferromagnet with electronic spins on the $b_1$ and $a_3$ sites pointing in the opposite directions. (c) Spin polarized half-metal resulting from the combination of Hubbard repulsion driven layer antiferromagnet and $D$-field induced layer polarization. The frozen spin orientation of the half-metal is arbitrary.            
}~\label{fig:RTLGHM}
\end{figure}

We begin by reviewing key experimental observations and summarizing our main results. Irrespective of the distance between the metallic gates, the Hubbard repulsion is expected to dominate in graphene-based layered materials at least near charge neutrality~\cite{katsnelson:hubbard}. It favors layer antiferromagnet, where electronic spins at the low-energy sites residing on the top and bottom layers point in the opposite directions [Fig.~\ref{fig:RTLGHM}(a),(b)]. By contrast, a $D$-field induces layer polarization of electronic density. Both orders lead to a uniform and isotropic gap in the quasiparticle spectra. But electronic bands loose two-fold spin degeneracy near each valley when they are present simultaneously, producing a spin-polarized, valley unpolarized \emph{half-metal}, when the chemical potential ($\mu$) lies between two bands [Fig.~\ref{fig:RTLGHM}(c)], as suggested by quantum oscillation measurements. Such Hubbard repulsion driven antiferromagnet, giving away to a half-metal in the presence of $D$-fields has also been reported in Ref.~\cite{Lau:RTLG}.  

\begin{figure}[t!]
\includegraphics[width=0.96\linewidth]{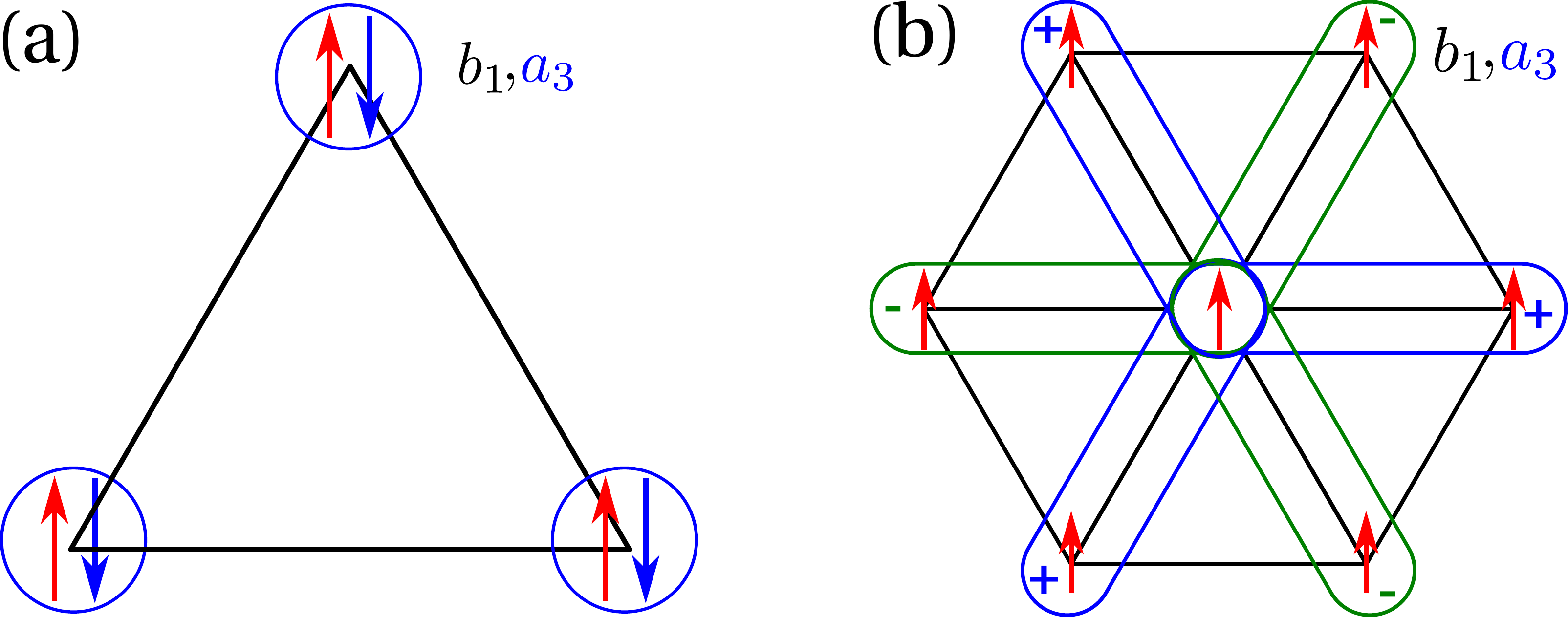}
\caption{Cooper pairs for (a) spin-singlet $s$-wave and (b) spin-triplet $f$-wave pairings, formed within the sites belonging to the same layer (top or bottom) of RTLG. The $s$-wave ($f$-wave) pairing is the candidate for SC1 (SC2)~\cite{young:RTLGSC}.      
}~\label{fig:swavefwavepairing}
\end{figure}

\begin{figure}[t!]
\includegraphics[width=0.96\linewidth]{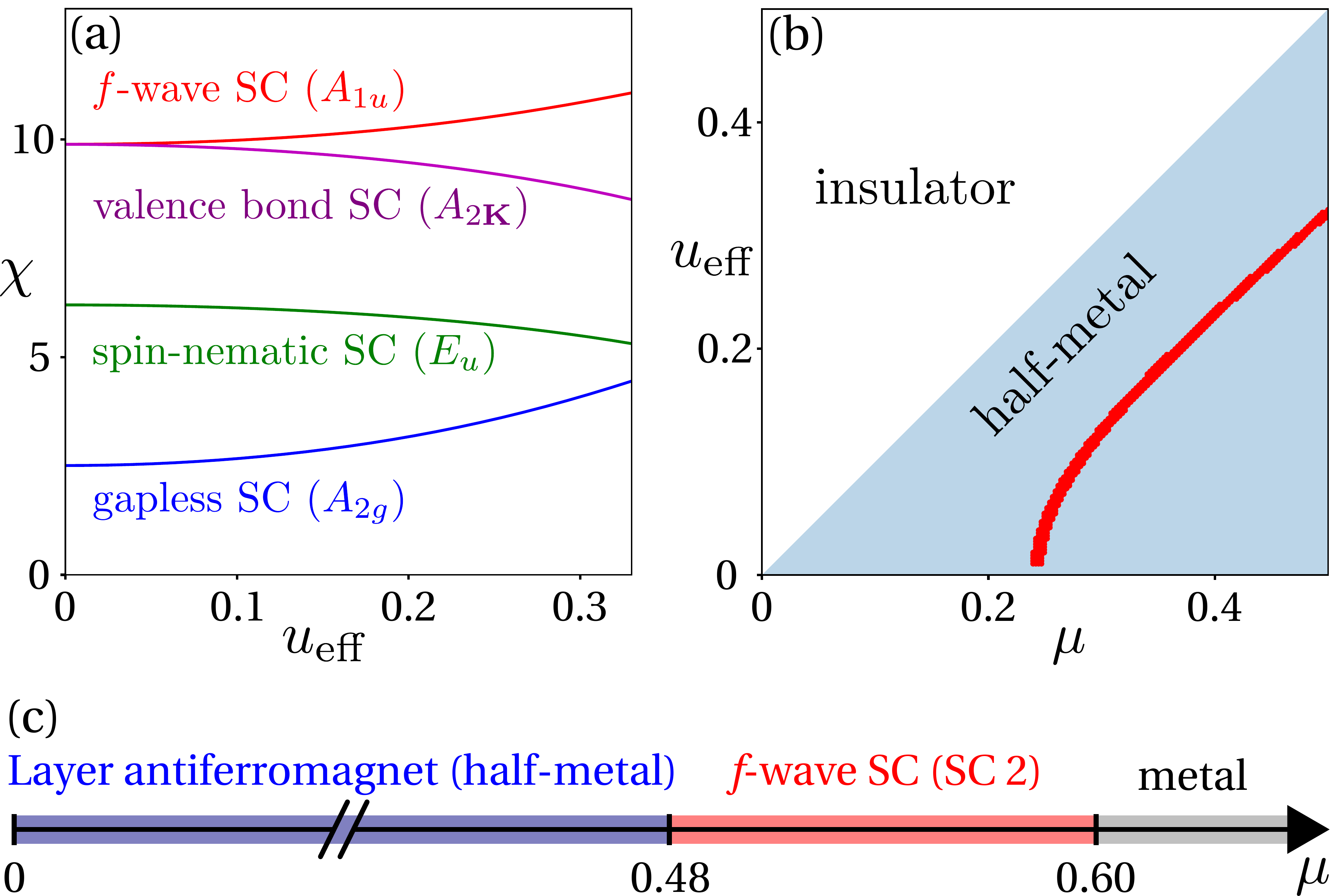}
\caption{(a) Mean field susceptibility $\chi$ [Eq.~(\ref{eq:susceptibility})] of spin-triplet pairings in a spin-polarized half-metal for $\mu=0.5$ and temperature $t=0.05$, where $u_{\rm eff}=|u-\Delta_{\rm LAF}|$. The $f$-wave pairing with the largest susceptibility is the prominent candidate for SC2. (b) The line of constant $f$-wave susceptibility ($\chi=20$) in the $(\mu, u_{\rm eff})$ plane qualitatively agrees with the observed line of superconductivity in the $(n_e,D)$ plane, where $n_e$ is the carrier density. (c) Half-metal and proximal $f$-wave pairing from repulsive interaction $\lambda_{A_{2u}}=0.43$ in the layer antiferromagnet channel for $u=0.6$ and $t=0.01$. Here $\chi$, $\mu$, $t$, $u_{\rm eff}$, $u$, $\lambda_{A_{2u}}$ are dimensionless~\cite{supplementary}.         
}~\label{fig:tripletsusceptibilityHM}
\end{figure}

The spin-polarized half-metal sustains only spin-triplet pairing among electrons with equal spin projection. With the assistance of the $D$-field an $f$-wave pairing [Fig.~\ref{fig:swavefwavepairing}(b)], producing isotropic gap on the Fermi surface of the half-metal, becomes energetically favored [Fig.~\ref{fig:tripletsusceptibilityHM}(a)]. The Cooper pairs are formed by electrons residing on the same layer. The superconducting wave-function changes sign six times under $2 \pi$ rotation in the real space and Brillouin zone. The phase diagram in the $(\mu, u)$ plane shows qualitative agreement with experiments [Fig.~\ref{fig:tripletsusceptibilityHM}(b)]. Here $\mu$ is measured from the cubic band touching point and $u$ is the voltage bias between the top and bottom layers, thus $D=-u/(2d_0)$, where $d_0$ is the interlayer separation. This triplet pairing (SC2) naturally exceeds the Pauli limiting in-plane magnetic field~\cite{young:RTLGSC}.

\begin{figure*}[t!]
\includegraphics[width=0.90\linewidth]{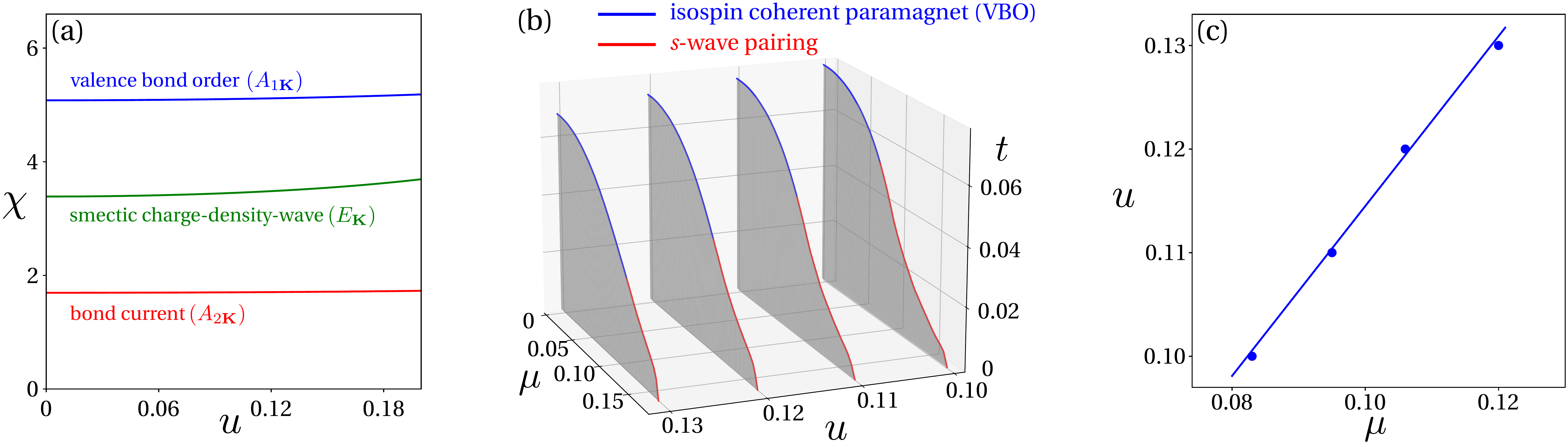}
\caption{(a)~Mean-field susceptibility ($\chi$) of isospin coherent paramagnets as a function of $u$ for $\mu=0.5$ and $t=0.05$, promoting VBO with largest susceptibility as its prominent candidate. (b) Various cuts of the phase diagram for a fixed bare interaction $\lambda{_{A_{1{\bf K}}}}=0.2$ showing a confluence of VBO and $s$-wave pairing. The shaded (white) region represents ordered (disordered) phase. (c) Onset points for $s$-wave pairing in the $(\mu,u)$ plane obtained from (b), showing qualitative similarities with experimental observation. For larger $\mu (u)$ the ordered phase is $s$-wave pairing (VBO). All the parameters are dimensionless~\cite{supplementary}.   
}~\label{fig:paramagnetswave}
\end{figure*}

As the system is tuned further away from charge neutrality, another superconducting state (SC1) appears from a valley coherent paramagnetic metal. In RTLG there are three candidates for such metallic state, among which the one featuring translational symmetry breaking valence bond order (VBO) of hopping between the low-energy sites is energetically superior [Fig.~\ref{fig:paramagnetswave}(a)]. Thus the valley coherent paramagnetic metal manifests VBO.

When the dominant repulsive electronic interaction is in the VBO channel, besides supporting the VBO order, it is also conducive for the condensation of spin-singlet $s$-wave pairing [Fig.~\ref{fig:paramagnetswave}(b)], as shown from a renormalization group (RG) calculation. Appearance of the $s$-wave pairing as a representative of SC1 is consistent with the observed Pauli limiting in-plane magnetic field~\cite{young:RTLGSC}. Weak $D$-field favoring such pairing is also in qualitative agreement with our findings [Fig.~\ref{fig:paramagnetswave}(c)]. Nucleation of the $s$-wave pairing follows a ``selection rule" among competing orders~\cite{royjuricic:selection, szabomoessnerroy:selection, szaboroy:selection}, as it constitutes a composite O(5) supervector of competing masses for cubic fermions in RTLG with VBO and induced layer polarization.

\emph{Model}.~We arrive at these results by considering the low-energy description for electronic bands in RTLG~\cite{RTLGband:1, RTLGband:2, RTLGband:3}. The unit cell is composed of six sites, each layer contributing two of them [Fig.~\ref{fig:RTLGHM}(a)]. But, one set of sites from the bottom ($a_1$) and top ($b_3$) layers reside on top of both set of sites ($b_2$ and $a_2$, respectively) of the middle layer (dimer-sites). The intralayer nearest-neighbor hopping ($t_0$) gives rise to massless Dirac fermions. But, the direct hopping ($t_\perp$) between the dimer-sites couples quasirelativistic Dirac fermions on honeycomb flatland as a static non-Abelian magnetic field~\cite{RTLGband:4} and pushes four out of six bands to high-energies ($t_\perp \approx 200$meV), the split-off bands, leaving only two bands near the charge neutrality point. They display cubic band touchings. Wavefunctions of the cubic bands predominantly live on the $b_1$ and $a_3$ sites near two inequivalent valleys. Accounting for the layer or equivalently sublattice, valley and spin degrees of freedom, we arrive at the single-particle Hamiltonian for noninteracting electrons 
\begin{equation}~\label{eq:hamilRTLG}
H_0= \alpha \left[ f_1 (\vec{k}) \Gamma_{3031} + f_2(\vec{k}) \Gamma_{3002} \right] + u \Gamma_{3003} -\mu \Gamma_{3000},
\end{equation}      
where $\alpha=t^3_0 a^3/t_\perp$, $a$ is the lattice spacing, $f_1(\vec{k})=k_x (k^2_x-3 k^2_y)$, $f_2(\vec{k})=-k_y(k^2_y-3 k^2_x)$. Momentum $\vec{k}$ is measured from the respective valleys. Electron (hole) doping corresponds to $\mu >0 \; (\mu<0)$. The sixteen-dimensional matrices are $\Gamma_{\mu \nu \rho \lambda}=\eta_\mu \sigma_\nu \tau_\rho \beta_\lambda$. Four sets of Pauli matrices $\{\eta_\mu \}$, $\{ \sigma_\nu \}$, $\{ \tau_\rho \}$ and $\{ \beta_\lambda \}$ operate on the particle-hole, spin, valley and layer indices, respectively, with $\mu, \nu, \rho, \lambda=0, \cdots, 3$. We introduce Nambu doubling (after a rotation by $\sigma_2$ and $\tau_1$ in the hole part~\cite{supplementary}) to facilitate the discussion on superconductivity. For $\mu=0$, the system is an insulator. Therefore, when $\mu$ reaches the bottom (top) of the conduction (valence) band, gated RTLG features discontinuity in density of states, known as the Van Hove singularity, at low doping.

We neglect particle-hole asymmetry and trigonal warping. The latter splits the momentum space vortex of vorticities $\pm 3$ near two valleys for cubic quasiparticles into three Dirac points with vorticities $\pm 1$, preserving the overall topology of the band touching points. As the density of states for cubic (Dirac) fermions diverges (vanishes) as $\varrho(E) \sim |E|^{-1/3}$ ($\varrho(E) \sim |E|$), strong correlation effects are driven by cubic fermions.

\emph{Half-metal}.~Near half-filling on site Hubbard repulsion favors layer antiferromagnet [Fig.~\ref{fig:RTLGHM}(b)], which when acquires a finite amplitude ($\Delta_{\rm LAF}$), is accompanied by the matrix $\Gamma_{0303}$. We choose the spin quantization axis in the $z$ direction. The energy spectra $\pm E_{\tau, \sigma}-\mu$, where $+(-)$ denotes the conduction (valence) band and 
\allowdisplaybreaks[4]
\begin{equation}
E_{\tau,\sigma} = \sqrt{\alpha^2 |\vec{k}|^6 + (u + \sigma \Delta_{\rm LAF})^2}, 
\end{equation} 
are valley degenerate (insensitive to $\tau=\pm$), but lack the spin degeneracy (for $\sigma=\pm$), since the matrices accompanying layer antiferromagnet ($\Gamma_{0303}$) and layer polarization ($\Gamma_{3003}$) commute with each other, and appear with the valley identity $\tau_0$ matrix. When $|u-\Delta_{\rm LAF}|<\mu<u+ \Delta_{\rm LAF}$, we realize a spin polarized, valley unpolarized half-metal [Fig.~\ref{fig:RTLGHM}(c)],  The effective single-particle Hamiltonian ($H^{\rm HM}_0$) for the half-metal takes the form of $H_0$ [Eq.~(\ref{eq:hamilRTLG})], with $u \to u_{\rm eff}= |u-\Delta_{\rm LAF}|$ and without the spin degrees of freedom, as the system behaves like an effective spinless one with frozen spin orientation.

The superconductor originating from the half-metal is constrained to be spin-triplet, with Cooper pairs forming between electrons with equal spin projection. RTLG altogether supports four spin-triplet superconductors: an $f$-wave pairing ($A_{1u}$), a pair-density-wave ($A_{2{\bf K}}$), a spin nematic pairing ($E_u$) and a gapless pairing ($A_{2g}$). Their irreducible representations under the $D_{3d}$ group are shown in the parentheses. The pairing matrices are  
\begin{equation}
\Gamma_{\mu 3 0}, \: (\Gamma_{\mu 12}, \Gamma_{\mu 22}), \: (\Gamma_{\mu 31}, \Gamma_{\mu 02}), \: \text{and} \: \Gamma_{\mu 33}, \: \text{respectively}. \nonumber 
\end{equation}      
Here $\Gamma_{\mu \rho \lambda}=\eta_\mu \tau_\rho \beta_\lambda$, and $\mu=1,2$ reflects the U(1) gauge redundancy of the superconducting phase.

To compare the propensities toward triplet paired states, we compute their bare mean field susceptibility for zero external frequency and momentum
\allowdisplaybreaks[4]	
\begin{eqnarray}~\label{eq:susceptibility}
\hspace{-0.5cm}\chi= -T \sum^{\infty}_{n=-\infty} \int \frac{d^2 \vec{k}}{(2\pi)^2} {\rm Tr} \left[ G(i \omega_n, \vec{k}) M G(i \omega_n, \vec{k}) M \right].
\end{eqnarray}
Here $M$ is the pairing matrix. The fermionic Green's function $G(i \omega_n, \vec{k})=[i \omega_n -H^{\rm HM}_0]^{-1}$, $\omega_n=(2 n+1) \pi T$ are the fermionic Matsubara frequencies, $T$ is the temperature, and the Boltzmann constant $k_{_B}=1$~\cite{supplementary}. As the $f$-wave pairing~\cite{honerkamp:fwave} and pair-density-wave~\cite{royherbut:kekule} are superconducting masses, they possess the largest degenerate susceptibility for zero $D$-field [Fig.~\ref{fig:tripletsusceptibilityHM}(a)].

External $D$-field lifts this degeneracy, as the $f$-wave (density-wave) pairing anticommutes (commutes) with $\Gamma_{303}=\eta_3 \tau_0 \beta_3$, the matrix accompanying the $D$-field in the half-metal manifold. In general, the $D$-field increases propensity toward any pairing that anticommutes with $\Gamma_{303}$. Still, the $f$-wave pairing possesses the largest susceptibility for any finite $\mu$ and $u_{\rm eff}$, standing as a promising candidate for the observed triplet pairing SC2. It can result from repulsive electronic interaction in the antiferromagnet ($\lambda_{A_{2u}}$) channel when the $D$-field is strong~\cite{supplementary} [Fig.~\ref{fig:tripletsusceptibilityHM}(c)] or electron-phonon interaction~\cite{dassarma:RTLG}.

As $D \propto u$ or $u_{\rm eff}$ and carrier density $n_e$ increases (decreases) with increasing (decreasing) $\mu$ in a half-metal, we can scrutinize the effect of $D$-field and $n_e$ on SC2 from the dependence of $\chi$ on $u_{\rm eff}$ and $\mu$. Since $\chi \propto T_c$ (transition temperature) and $\Delta$ (amplitude) of the paired state, a line of constant $\chi$ in the $(\mu, u_{\rm eff})$ plane should qualitatively mimic the observed line of superconductivity in the $(n_e, D)$ plane. With increasing $D$-field layer polarization increases, bringing the system closer to insulation. To compensate such field induced insulation, a larger $n_e$ or $\mu$ is required to induce superconductivity, as found in experiments and also in our scaling of constant $\chi$ line in the $(\mu, u_{\rm eff})$ plane [Fig.~\ref{fig:tripletsusceptibilityHM}(b)].

\emph{Paramagnetic metal}.~We now proceed to larger doping regime, where the system supports an isospin coherent paramagnetic metal, bordering a superconductor (SC1)~\cite{young:RTLGSC}. Isospin coherence implies that two valleys get coupled, resulting in breaking of translational symmetry. In the low-energy continuum limit it translates into a U(1) symmetry, generated by $\Gamma_{0030}=\eta_0 \sigma_0 \tau_3 \beta_0$~\cite{szaboroy:selection, HJR2009}. The paramagnetic nature of the metallic state indicates that it is spin-singlet, for which there are three candidates: VBO ($A_{1{\bf K}}$), bond current ($A_{2{\bf K}}$) and smectic charge-density-wave ($E_{\bf K}$), accompanied by the matrices 
\allowdisplaybreaks[4]	
\begin{equation}
\Gamma_{30 \rho 1}, \Gamma_{00 \rho 2},\: \text{and} \: (\Gamma_{30 \rho 0}, \Gamma_{30 \rho 3}), \: \text{respectively}. \nonumber  
\end{equation} 
Their irreducible representations in the $D_{3d}$ group are shown inside the parentheses. Here $\rho=1,2$ manifests the U(1) valley coherence. The smectic charge-density-wave also breaks rotational symmetry about the $z$ direction, generated by $\Gamma_{0033}$. We compute the susceptibilities of these orders at finite $\mu$ and $u$ [Eq.~(\ref{eq:susceptibility})] with $G(i \omega_n, \vec{k})=[i \omega_n -H_0]^{-1}$~\cite{supplementary}. The VBO possesses the largest susceptibility and is thus the most prominent candidate for the isospin coherent paramagnetic metal [Fig.~\ref{fig:paramagnetswave}(a)]. This is so because VBO is a mass for cubic quasiparticles and the associated matrices ($\Gamma_{3011}, \Gamma_{3021}$) also anticommute with the one accompanying $u$ ($\Gamma_{3003}$).

To shed light on the nature of the pairing, originating from such a correlated metal, we perform a leading-order RG analysis with the four-fermion interaction 
\begin{equation}
g_{_{A_{1{\bf K}}}} \left[ \left( \Psi^\dagger \Gamma_{3011} \Psi \right)^2 + \left( \Psi^\dagger \Gamma_{3021} \Psi \right)^2 \right] \nonumber  
\end{equation}
that favors VBO. Sixteen-component Nambu-doubled spinors $\Psi^\dagger$ and $\Psi$ involve spin, valley and layer degrees of freedom~\cite{supplementary}. Besides demonstrating the stability of VBO, we showcase the emergence of superconductivity at low temperatures and finite $\mu$ from a pure \emph{repulsive} interaction $g_{_{A_{1{\bf K}}}}>0$, following the Kohn-Luttinger mechanism~\cite{kohnluttinger, chubukov:review}. We integrate out the fast Fourier modes from a Wilsonian shell $\Lambda e^{-\ell} < |\vec{k}|< \Lambda$. Here $\Lambda$ is the ultraviolet momentum cutoff up to which the quasiparticle spectra remain cubic and $\ell$ is the logarithm of the RG scale. The coupled RG flow equations read 
\begin{eqnarray}
\frac{d \lambda{_{A_{1{\bf K}}}}}{d \ell}= \lambda{_{A_{1{\bf K}}}} + \lambda{^2_{A_{1{\bf K}}}} H(t, \mu, u) 
\:\:\: \text{and} \:\:\:
\frac{d x}{d \ell}= 3 x,\label{eq:beta_interaction}
\end{eqnarray}    
for $x=t,\mu, u$. The dimensionless quantities are $\lambda{_{A_{1{\bf K}}}}=g{_{_{A_{1{\bf K}}}}}/(2 \pi \alpha \Lambda)$, $t=T/(\alpha \Lambda^3)$, $\tilde{\mu}=\mu/(\alpha \Lambda^3)$ and $\tilde{u}=u/(\alpha \Lambda^3)$. For brevity we take $\tilde{\mu} \to \mu$ and $\tilde{u} \to u$. The $H$ function is shown in the Supplementary Materials (SM)~\cite{supplementary}. Even though $\lambda{_{A_{1{\bf K}}}}$ is a \emph{relevant} parameter due to the divergent density of states of cubic fermions, its RG flow terminates at an infrared scale $\ell^\star={\rm min}. (\ell^\star_t, \ell^\star_\mu, \ell^\star_u)$, where $\ell^\star_x= \ln [x^{-1}(0)]/3$, with $x(0)<1$ as its bare value. Then the system describes an ordered (a disordered) phase when $\lambda{_{A_{1{\bf K}}}}(\ell^\star) >1 \; (<1)$.

Only VBO and spin-singlet $s$-wave pairing can be realized in the ordered phase. To capture their competition, we simultaneously allow the conjugate fields, coupling with the corresponding fermion bilinears as  
\begin{equation}
\Delta_{A_{1{\bf K}}} \sum_{\rho=1,2} \Psi^\dagger \Gamma_{30 \rho 1} \Psi 
\:\: \text{and} \:\:
\Delta_{A_{1g}} \sum_{\mu=1,2} \Psi^\dagger \Gamma_{\mu 0 0 0} \Psi,  \nonumber 
\end{equation}
respectively, to flow under coarse grain, leading to
\begin{equation}
\frac{d \ln \Delta_{y}}{d \ell} -3= \lambda{_{A_{1{\bf K}}}} I_y (t, \mu, u), \:\: 
\text{for} \:\: y=A_{1{\bf K}} \:\: \text{and} \:\: A_{1g}. \label{eq:beta_conjField}
\end{equation}       
The $I$ functions are shown in the SM~\cite{supplementary}. As $\lambda{_{A_{1{\bf K}}}}$ diverges, indicating onset of an ordered phase, the pattern of symmetry breaking is set by the conjugate field that diverges toward $+\infty$ fastest. Following this prescription we construct few cuts of the phase diagram in the $(\mu, t)$ plane for a fixed bare $\lambda{_{A_{1{\bf K}}}}$ and for various $u$ [Fig.~\ref{fig:paramagnetswave}(b)].

At sufficiently low (high) temperatures the ordered phase supports an $s$-wave pairing (a VBO metal), manifesting an ``organizing principle" based on a generalized energy-entropy argument~\cite{szabomoessnerroy:selection, szaboroy:selection}. The $s$-wave superconductor isotropically gaps the entire Fermi surface and is thus energetically favored at low temperature. The VBO metal occupies the high-temperature regime, as it carries more entropy due to the presence of gapless quasiparticles. The paired state stems from quantum fluctuations of incipient VBO without any long range order. With increasing $u$ or $D$-field as the system appears at the shore of insulation, a larger chemical doping is required to induce superconductivity [Fig.~\ref{fig:paramagnetswave}(c)], similar to the situation for SC2 and as also observed in experiments. The appearance of the $s$-wave pairing as SC1 follows a ``selection rule"~\cite{royjuricic:selection, szabomoessnerroy:selection, szaboroy:selection}, since the matrices involving the paired state, VBO and $D$-field induced layer polarization
\begin{equation}
\{ \Gamma_{1000}, \Gamma_{2000}, \Gamma_{3011}, \Gamma_{3021}, \Gamma_{3003}\} \nonumber
\end{equation}  
constitute a unique O(5) composite vector of competing masses for cubic fermions.

\emph{Quarter-metal}.~The mechanism for the half-metal can now be extended to predict the nature of the observed quarter-metal~\cite{young:RTLGMetal}. Notice that with the onset of the quantum anomalous Hall order, which can be triggered by intralayer next-nearest-neighbor repulsion ($V_2$) the residual two-fold valley degeneracy of a half-metal gets lifted and the system supports a spin- and valley-polarized quarter-metal at low doping~\cite{supplementary}. This is so because the matrix accompanying the anomalous Hall order ($\Gamma_{0033}$), commutes with the matrices for the layer antiferromagnet ($\Gamma_{0303}$) and layer polarization ($\Gamma_{3003}$). This prediction has been supported by the observed hysteresis in the presence of weak perpendicular magnetic field, stemming from the orbital magnetism of anomalous Hall order. In RTLG this phase supports intralayer circulating currents among the next-nearest-neighbor sites belonging to the $b_1$ and $a_3$ sublattices, but in the opposite directions.

\emph{Discussions}.~Guided by recent experiments~\cite{young:RTLGSC, young:RTLGMetal}, here we identify the prominent candidates of half- and quarter-metals and their proximal superconductors. Only the $A_{2{\bf K}}$ pair-density-wave featuring spin- and valley-polarized Cooper pairs can be realized in a quarter-metal, which remains to be observed. The VBO should feature new diffraction peaks in a paramagnetic metal due to broken translational symmetry. The pairing nature can be pinned from tunneling spectroscopy (gapped or gapless) and Josephson junction (pairing symmetry). Both half- and quarter-metals have recently been observed in Bernal bilayer graphene~\cite{BBLGExp:1, BBLGExp:2, BBLGExp:3}, where identical mechanism remains operative~\cite{szaboroy:BBLG}. Our discussion should stimulate future experiments, underpinning the symmetries of these correlated phases.

\emph{Acknowledgments}.~B.R. was supported by a Startup grant from Lehigh University. We thank Andrea F. Young for useful correspondence.  

\emph{Note added}.~After completing this work, we became aware of few preprints, also discussing competing phases and superconductivity in RTLG~\cite{RTLGrecent:1, RTLGrecent:2, RTLGrecent:3}.


\end{document}